\documentclass[%
preprint,
nofootinbib,
amsmath,amssymb,
pre,
]{revtex4-2}

\usepackage{graphicx}
\usepackage{dcolumn}
\usepackage{bm}

\usepackage{amsmath}
\usepackage{amssymb}
\usepackage{amsthm}
\usepackage{url}
\usepackage{booktabs}
\usepackage{hhline}
\usepackage{comment}
\usepackage{array}
\usepackage[table]{xcolor}

\begin{document}


\title{Simple control for complex pandemics}

\author{Sarah C.~Fay}
\email{Corresponding author: scfay@mit.edu}
 \affiliation{Institute for Data Systems \& Society \\ Massachusetts Institute of Technology \\ Cambridge, MA 02139, USA}%
 
\author{Dalton J.~Jones}%
\affiliation{Institute for Data Systems \& Society \\ Massachusetts Institute of Technology \\ Cambridge, MA 02139, USA}%

\author{Munther A.~Dahleh}
\affiliation{Institute for Data Systems \& Society \\ Massachusetts Institute of Technology \\ Cambridge, MA 02139, USA}%

\author{A.E.~Hosoi}
\affiliation{Institute for Data Systems \& Society \\ Massachusetts Institute of Technology \\ Cambridge, MA 02139, USA}%

\date{\today}


\begin{abstract}
The COVID-19 pandemic began over two years ago, yet schools, businesses, and other organizations are still struggling to keep the risk of disease outbreak low while returning to (near) normal functionality. Observations from these past years suggest that this goal can be achieved through the right balance of mitigation strategies, which may include some combination of mask use, vaccinations, viral testing, and contact tracing. The choice of mitigation measures will be uniquely based on the needs and available resources of each organization. This article presents practical guidance for creating these policies based on an analytical model of disease spread that captures the combined effects of each of these interventions. The resulting guidance is tested through simulation across a wide range of parameters and used to discuss the spread of disease on college campuses. 
\end{abstract}
\maketitle
\section{INTRODUCTION}
Communities around the world are fighting the COVID-19 pandemic. In the United States alone, more than 50 million cases and 800,000 deaths have been recorded \cite{CDCCOVID}. The most reliable way to end transmission of this and other rapidly spreading diseases is to stop any and all contact between individuals in the population; no interaction means no transmission. While this strategy is extreme, many communities (even nations) implemented some version of this plan in the early weeks and months of the COVID-19 pandemic. These guidelines and mandates for social separation (distancing) were largely successful in slowing the spread of the disease \cite{socialDistancing1, socialDistancing2, socialDistancing3}, but they also prevented healthy people from interacting with each other, hindering global economic activity \cite{econ1, econ2} and leaving many feeling isolated and lonely \cite{lonely1, lonely2}. 

Reopening schools, businesses and other organizations and returning to a more normal level of interaction is therefore a priority for many communities. However, this does not mean that we must expect and accept community-wide outbreaks; other interventions can be implemented to limit the risk of transmission. Such interventions include (but are not limited to) mask use, vaccinations, viral testing, and contact tracing. This paper quantifies the effect of each of these mitigation strategies on the spread of disease and presents practical guidance for creating policies to maintain low risk of disease outbreak. The analysis that leads to this guidance is performed on a stochastic compartmental model for disease transmission, and the result is a single algebraic criterion (an effective disease reproduction number) based on the implemented interventions. In simulations of infections on random networks (agent-based simulations), this single criterion is able to predict the presence or absence of outbreaks across a vast array of population and disease parameters, whose true values will be unknown to those implementing intervention strategies. The practical guidance produced here was also the foundation of the policies regarding interventions implemented at the Massachusetts Institute of Technology (MIT), which has gradually resumed its normal functioning over the last two years without widespread outbreaks. Policies at other colleges and universities can also be considered in the framework of this analysis, and the outbreak trends at these other institutions are consistent with the analysis presented below given their chosen disease mitigation policies.

\section{METHODS}
\subsection{Modeling foundations}
We begin by applying the principles established by Kermack and McKendrick \cite{SIR_OG} for modeling the spread of infection. This model relies on partitioning the population into different categories (e.g.~Susceptible, Infectious, Recovered or SIR) based on disease status and applying a conservation law on the total population size. The rate of change of the number of people in a particular partition is determined only by the net flow of individuals into or out of that group. Stable dynamics (i.e.~a shrinking infectious population) can be equivalently defined as a positive net flow of people out of the infectious compartment.

\begin{figure*}
\centering
\includegraphics[width=0.9\linewidth]{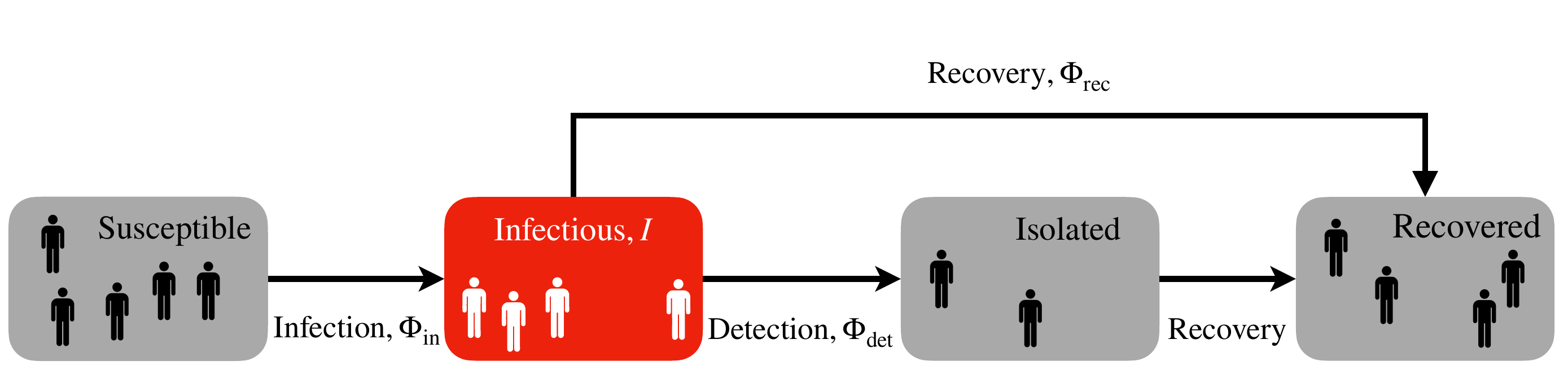}
\caption{Compartmental model for COVID-19 and similar diseases. Individuals join the infectious class (red) by contracting the disease (at a rate $\Phi_{\mathrm{in}}$) and leave by recovering ($\Phi_{\mathrm{rec}}$) or by being detected and subsequently isolated ($\Phi_{\mathrm{det}}$).}
\label{fig:CVdiagram}
\end{figure*}

Many extensions of the SIR model include additional classifications of individuals \cite{SIQRmodel, SEIRModel2, berger2020seir, ageModel}. In our cases, we partition the population into four groups: susceptible, infectious, isolated, and recovered.\footnote{An exposed/latent group, which separates presymptomatic from symptomatic individuals, is not considered in this study due to the high number of COVID-19 cases that never become symptomatic \cite{COVIDAsymptomatic, asympSpread}. Additionally, by ignoring the development of symptoms (and thus detection), we will end up \textit{overestimating} the number of total undetected infections, leading to more conservative policies to prevent outbreaks.} The choice of partitions and the flows between the compartments are depicted in Fig.~\ref{fig:CVdiagram}. The rate of the flow into the infectious class is denoted as $\Phi_{\mathrm{in}}$. The flow out is driven by two mechanisms: recovery ($\Phi_{\mathrm{rec}}$) and detection ($\Phi_{\mathrm{det}}$).

We can now write a conservation equation for the infectious group:
\begin{equation}
   I_{t+1} - I_t = \Phi_{\mathrm{in}} - \Phi_{\mathrm{rec}} - \Phi_{\mathrm{det}}
    \label{eq:CVeqn}
\end{equation}
where $I_t$ is the number of infectious individuals at time $t$. This conservation equation highlights the three levers that we can operate to reduce the spread of the disease and achieve controlled, stable dynamics. The first is reducing the rate of infection ($\Phi_{\mathrm{in}}$), which can be achieved by increasing social distancing, wearing masks, or vaccinating the population. The second is increasing the rate of recovery. This lever is not easy to operate and is dictated by biological constraints and available medical treatments. The final lever, increasing detection, is achieved by ramping up infection testing and contact tracing measures. 

The objective of this analysis is to provide practical guidance to organizations looking to reopen. For this reason, the model must capture the fact that, in practice, compliance to these intervention strategies may not be perfect; organizations may not be able to mandate particular mitigation strategies and even if they do, individuals may choose not to abide. In order to capture the effect (and danger) of this non-compliance, the model presented here allows for some members of the population to employ the intervention strategies while others abstain. The fraction of the population who are vaccinated $f_v$, who opt into random surveillance testing $f_T$, and who wear masks $f_m$ are therefore all considered. While a person's decision to get vaccinated or to opt into a surveillance testing program is fixed, their decision to wear a mask may vary day-to-day. Such considerations are made in formalizing the partitions and flows between partitions of the model.

Stability in the disease dynamics is defined as the case where the size of the infectious population (shown in red in Fig.~\ref{fig:CVdiagram}) is shrinking, meaning this is the partition we are most concerned with modeling. For this study, we will assume that vaccinated individuals and unvaccinated individuals will behave the same way when they are infectious, i.e.~they will have the same ability to transmit the disease and will be equally likely to recover or be tested/detected on a given day. Additionally, because we do not assume masking behavior is constant day-to-day, we need not separate the infectious population into masked and unmasked partitions. Opting into or out of a surveillance testing program, however, is a different story. An individual's testing status (their opt-in or opt-out decision) remains the same over time and infected individuals who are tested will become isolated and leave the infectious population much faster than those who opt-out of testing. Therefore, we need to consider the behavior of the opt-into testing population and the opt-out-of testing population separately, meaning we will examine two conservation principles, one for those opting into testing
\begin{equation}
   I_{t+1,\mathrm{oi}} - I_{t, \mathrm{oi}} = \Phi_{\mathrm{in}, \mathrm{oi}} - \Phi_{\mathrm{rec},\mathrm{oi}} - \Phi_{\mathrm{det},\mathrm{oi}}
   \label{eq:CVeqnOI}
\end{equation}
and one for those opting out
\begin{equation}
   I_{t+1,\mathrm{oo}} - I_{t, \mathrm{oo}} = \Phi_{\mathrm{in}, \mathrm{oo}} - \Phi_{\mathrm{rec},\mathrm{oo}} - \Phi_{\mathrm{det},\mathrm{oo}}.
   \label{eq:CVeqnOO}
\end{equation}

The next step is defining these flows into and out of the infectious populations. First, we consider the infection process. Infection occurs when an infectious individual successfully transmits the disease to one of their susceptible contacts. To estimate the likelihood of transmission, we must first introduce the basic reproduction number of a disease, denoted $\mathcal{R}_0$. This classic epidemiological parameter is defined as the number of secondary cases resulting from a single infected individual in a fully susceptible population \cite{IntroMathEpid, SIR_OG}. If, for instance, that single infected individual was contagious for $d$ days, had $\hat{x}$ daily contacts (all susceptible individuals), and there was a likelihood of transmitting the disease to each contact (each day) of $\rho$, the basic reproduction number would be
\begin{equation}
    \mathcal{R}_0 = \rho \hat{x} d.
    \label{eq:R0}
\end{equation}
This relationship defines the transmission likelihood in the absence of mitigation strategies ($\rho$) in terms of the parameters most often reported for a particular infectious disease, i.e.~$\mathcal{R}_0$ and $d$. The daily number of susceptible contacts $\hat{x}$ will depend on the particular community and will be discussed further below.

The transmission likelihood in a given interaction, however, will be reduced when interventions such as mask-wearing and vaccination are implemented. If one person is wearing a mask, the likelihood of transmission is reduced by a factor of $\left(1-\epsilon_m\right)$ where $\epsilon_m$ is the efficacy of mask use at preventing the spread of the disease. If both people in the interaction are wearing masks, the reduction factor is $\left(1-\epsilon_m\right)^2$. Similarly, if the susceptible individual is vaccinated, the transmission likelihood is reduced by a factor of $\left(1-\epsilon_v\right)$ with vaccine efficacy to breakthrough infectious $\epsilon_v$. If the susceptible individual is a testing opt-in individual, they will join the testing opt-in infectious population when they contract the disease. If instead the susceptible individual is a testing opt-out individual, they will join the opt-out infectious population when they contract the disease. 

All together, these parameters can be used to define the flow of susceptible individuals into the infectious compartments:
\begin{equation}
\begin{split}
    \Phi_{\mathrm{in},oi} &= \sum_{j=1}^{I_{t,\mathrm{oi}}}\sum_{k=1}^{X_j} T_k Y_{jk} + \sum_{j=1}^{I_{t,\mathrm{oo}}}\sum_{k=1}^{X_j} T_k Y_{jk} \\
    \Phi_{\mathrm{in},oo} &= \sum_{j=1}^{I_{t, \mathrm{oi}}}\sum_{k=1}^{X_j} \left(1-T_k\right) Y_{jk} + \sum_{j=1}^{I_{t, \mathrm{oo}}}\sum_{k=1}^{X_j} \left(1-T_k\right) Y_{jk} 
\end{split}
\end{equation}
where $X_j$ is a random variable representing the number of susceptible contacts of infected individual $j$. For now, we will assume we know this distribution and that it has a mean $\mu_x$ and a variance $\sigma^2_x$, though we will find that in determining stability to outbreaks, the basic reproduction number $\mathcal{R}_0$ is sufficient for capturing the critical aspects of this distribution. The testing status (opt-in or opt-out) of the susceptible contact $k$ is captured by the Bernoulli random variable $T_k$ which has parameter $f_T$ and indicates whether or not that individual will join the opt-in or opt-out population upon infection.

The random variable $Y_{jk}$ is a Bernoulli random variable with parameter equal to the effective transmission likelihood
\begin{equation}
    \rho_{\mathrm{eff}} = \rho \left(1-\epsilon_m M_j\right)\left(1- \epsilon_m M_k \right)\left( 1- \epsilon_v V_k \right)
    \label{eq:rhoEff}
\end{equation}
which itself depends on the normal (no intervention) transmission probability $\rho$, the intervention efficacies $\epsilon_m$ and $\epsilon_v$, and the intervention statuses of both individuals determined by the Bernoulli random variables $M_j$ (mask status of $j$, parameter $f_m$), $M_k$ (mask status of $k$, parameter $f_m$), and $V_k$ (vaccine status of $k$, parameter $f_v$).

Next we consider the recovery process. A simple approximation is to assume that each infected individual is equally likely to recover on any given day and that the expected duration for which they are infectious (in the absence of interventions) is $d$. The resulting recovery flows are
\begin{equation}
\begin{split}
    \Phi_{\mathrm{det}, \mathrm{oi}} &= \sum_{j=1}^{I_{t, \mathrm{oi}}} Z_{j}\\
    \Phi_{\mathrm{det}, \mathrm{oo}} &= \sum_{j=1}^{I_{t, \mathrm{oo}}} Z_{j}
\end{split}
\end{equation}
where $Z_j$ is a Bernoulli random variable with parameter $1/d$. 

Finally, we consider detection through infection testing and follow-up contact tracing. While only opt-in individuals will be detected through random surveillance testing, both their opt-in and opt-out contacts may be identified and isolated through follow-up contact tracing. In the surveillance (random) testing group, any member is equally likely to be selected for testing on a given day (with probability $\nu$, the daily testing frequency). If an individual tests positive, their contacts are traced and tested. Isolating contacts only has a substantial effect on disease dynamics when those contacts themselves are infectious at the time of tracing. At the previous time step, each contact had a likelihood $\rho_{\mathrm{eff}}$ of having been infected by person $j$. Using this as the likelihood that contact $k$ is infected at the current time step will \textit{underestimate} the number of infectious contacts (because person $j$ may have been infecting their contacts for more than one day), which will lead to more conservative stability bounds. We also consider that not all contacts will be successfully traced, an event which happens with likelihood $\epsilon_c$. Each contact therefore has a probability $\epsilon_c \rho_{\mathrm{eff}}$ of being detected as infectious through contact tracing. Together, the number of infectious individuals removed from the infectious groups by detection via both random testing and contact tracing are
\begin{equation}
\begin{split}
    \Phi_{\mathrm{det}, \mathrm{oi}} &= \sum_{j=1}^{I_{t,\mathrm{oi}}} V_j \left(1 + \sum_{k=1}^{X_j} T_k W_{jk} \right) \\
    \Phi_{\mathrm{det}, \mathrm{oo}} &= \sum_{j=1}^{I_{t,\mathrm{oi}}} V_j \left(0+ \sum_{k=1}^{X_j} \left(1-T_k\right)W_{jk}  \right)
\end{split}
\end{equation}
where $V_j$ (random testing) and $W_{jk}$ (tracing) are Bernoulli random variables with parameters $\nu$ and $\epsilon_c \rho_{\mathrm{eff}}$ respectively. Again, the testing status is captured in the random variable $T_k$ as defined above and the transmission probability $\rho_{\mathrm{eff}}$ is affected by the mask and vaccine statuses of the individuals involved (see Eq.~\ref{eq:rhoEff}).

Using the the constitutive relationships defined above and the conservation laws of Eq.~\ref{eq:CVeqnOI}-\ref{eq:CVeqnOO}, we can rewrite the equations for the evolution of the number of infectious individuals in each testing group:
\begin{equation}
\begin{split}
    I_{t+1, \mathrm{oi}} &= I_{t, \mathrm{oi}} + \sum_{j=1}^{I_{t, \mathrm{oi}}} \left[\left(\sum_{k=1}^{X_j} T_k \left(Y_{jk} - V_j W_{jk}\right)\right) - Z_j \left(1-V_j\right) - V_j \right] + \sum_{j=1}^{I_{t,\mathrm{oo}}}\left[ \sum_{k=1}^{X_j} T_k Y_{jk} \right] \\
    I_{t+1, \mathrm{oo}} &= I_{t, \mathrm{oo}} + \sum_{j=1}^{I_{t, \mathrm{oi}}} \left[\sum_{k=1}^{X_j} \left(1-T_k\right) \left(Y_{jk} - V_j W_{jk}\right)\right] + \sum_{j=1}^{I_{t,\mathrm{oo}}}\left[\left( \sum_{k=1}^{X_j} \left(1-T_k\right) Y_{jk} \right) - Z_j\right]. 
\end{split}
\label{eq:stochasticDiff}
\end{equation}
Here, we have made one additional correction. The recovery removal for each member of the opt-in population is $Z_j\left(1-V_j\right)$ to avoid double counting people who recover on the same time step that they test positive. This ensures we are not overestimating the flow out of the infectious population, hence our results will remain conservative.

\subsection{Stability analysis}
The disease dynamics in a community are considered stable if a randomly infected individual will not be able to spread the disease to others in the community before being detected or recovering. In a real (stochastic) environment it will always be possible for a single infection to lead to a widespread outbreak, as in the case of ``superspreader" events where a single infectious individual has many more contacts than usual on a single day and will likely be able to spread the disease before being identified. To capture the dynamics, therefore, we consider the mean and variance of the evolution of the number of infectious individuals at a given time. 

The mean dynamics are found by taking the expectation of Eq.~\ref{eq:stochasticDiff} for the full stochastic process. Doing so leads us to find a 2-by-2 transition matrix: 
\begin{equation}
    \textbf{A} = 
    \left[\begin{array}{cc}
        a_{11} & a_{12} \\
        a_{21} & a_{22}
    \end{array} \right] 
    =
    \left[\begin{array}{cc}
        1+\hat{\rho}_{\mathrm{eff}}\mu_x f_T \left(1-\nu \epsilon_c\right) - \left(\frac{1}{d} + \nu - \frac{\nu}{d}\right)& \hat{\rho}_{\mathrm{eff}} \mu_x f_T \\
        \hat{\rho}_{\mathrm{eff}}\mu_x \left(1-f_T\right)\left(1-\nu \epsilon_c\right) & 1+ \hat{\rho}_{\mathrm{eff}} \mu_x \left(1-f_T\right) - \frac{1}{d}
    \end{array} \right] 
\end{equation}
such that
\begin{equation}
    \left(\begin{array}{cc}
         \mathbb{E}\left[ I_{t+1, \mathrm{oi}}\right] \\
         \mathbb{E} \left[I_{t+1, \mathrm{oo}}\right] 
    \end{array}\right) = 
    \textbf{A}
    \left(\begin{array}{cc}
         \mathbb{E} \left[I_{t, \mathrm{oi}} \right]\\
         \mathbb{E} \left[I_{t, \mathrm{oo}}\right] 
    \end{array} \right).
    \label{eq:2by2}
\end{equation}
Here, we have used 
\begin{equation}
    \hat{\rho}_{\mathrm{eff}} = \mathbb{E}\left[\rho_{\mathrm{eff}}\right] = \rho \left(1- \epsilon_m f_m\right)^2 \left(1-\epsilon_v f_v\right).
\end{equation}

Stability in the mean dynamics therefore is achieved when the eigenvalues of $\textbf{A}$, the transition matrix, are both less than 1. Finding the eigenvalues of the matrix in Eq.~\ref{eq:2by2} leads to a single stability criterion; this system is stable if
\begin{equation}
    1 > \rho \mu_x d \left(1-\epsilon_m f_m\right)^2 \left(1- \epsilon_v f_v\right)\left(f_T \frac{1-\nu \epsilon_c}{1+\nu\left(d-1\right)} + \left(1-f_T\right)\right)
    \label{eq:stability}
\end{equation}

Notice, the mean value of the distribution $X$ (the number of susceptible contacts of an infectious individual) appears in this expression, implying that we need to know something about the underlying distribution, information that is typically unavailable. In analyzing stability to create practical guidance for intervention measures, this distribution (particularly its mean) need not be known exactly, but we do need an upper bound on this value to ensure that we overestimate disease transmission and determine conservative values for mitigation requirements. To estimate this upper bound, we return to the discussion of the basic reproduction number $\mathcal{R}_0$ (see Eq.~\ref{eq:R0}). Certainly, assuming that all of an individual's contacts are susceptible (which is implied in the $\hat{x}$ of the basic reproduction number) is an overestimate of the true number of susceptible contacts that an individual has. 

However, we are not considering the number of susceptible contacts that any random individual in the population has; we are considering the number of susceptible contacts that an \textit{infected} individual has, which, akin to the friendship paradox, is expected to be higher than that of non-infected individuals \cite{KojakuSadamori2021_friendship}. In this case, $\hat{x}$ is still a reasonable (over)estimate for $\mathbb{E}\left[X\right]$ because it is inferred from the measured $\mathcal{R}_0$ (based on infection case data), which naturally incorporates the tendency of a disease to spread to highly-connected individuals \cite{HeffernanJM2005_measureR0, KEELINGMATTJ2000_measureR0, DelamaterPaulL2019ComplexR0, LiJing2011T_failureR0}. In conclusion, the term $\rho\mu_x$, which contains parameters that are unknown and difficult to measure, can be rewritten as $\mathcal{R}_0/d$ without jeopardizing confidence in the resulting stability bound.

Now we return to Eq.~\ref{eq:stability} to redefine it in more familiar terms, i.e.~in terms of an effective reproduction number $\mathcal{R}_{\mathrm{eff}}$:
\begin{equation}
    \mathcal{R}_{\mathrm{eff}} = \mathcal{R}_0 \left(1-\epsilon_m f_m\right)^2 \left(1- \epsilon_v f_v\right)\left(f_T \frac{1-\nu \epsilon_c}{1+\nu\left(d-1\right)} + \left(1-f_T\right)\right) < 1
    \label{eq:ReffStability}
\end{equation}
which ensures stability of the system. The implications of this expression, i.e.~how it can be used to quantify trade-offs between mitigation strategies, are discussed in the following sections. One important note, however, is that this expression depends on the underlying network of the community only through the basic reproduction number $\mathcal{R}_0$, a commonly measured and reported value for a disease. No further information about the network, such as the variance in the distribution of contacts $X$, is required to determine a sufficient balance of interventions to ensure stability.

Having explored the evolution of the mean number of infectious individuals, we now turn to the evolution of the variance. All of these dynamic equations together can be written as
\begin{equation}
    \left(\begin{array}{cc}
         \mathbb{E}\left[ I_{t+1, \mathrm{oi}}\right] \\
         \mathbb{E} \left[I_{t+1, \mathrm{oo}}\right] \\
         \mathrm{var} \left(I_{t+1, \mathrm{oi}} \right)\\
         \mathrm{var} \left(I_{t+1, \mathrm{oo}} \right)
    \end{array}\right) = 
    \left[\begin{array}{cccc}
        a_{11} & a_{12} & 0 & 0 \\
        a_{21} & a_{22} & 0 & 0 \\
        b_{11} & b_{12} & a_{11}^2 & a_{12}^2 \\
        b_{21} & b_{22} & a_{21}^2 & a_{22}^2 
    \end{array} \right] 
    \left(\begin{array}{cc}
         \mathbb{E} \left[I_{t, \mathrm{oi}} \right]\\
         \mathbb{E} \left[I_{t, \mathrm{oo}}\right] \\
         \mathrm{var} \left(I_{t, \mathrm{oi}} \right)\\
         \mathrm{var} \left(I_{t, \mathrm{oo}} \right)
    \end{array} \right).
    \label{eq:4by4}
\end{equation}
where the elements $b_{11}, b_{12}, b_{21},$ and $b_{22}$ are functions of the disease and mitigation parameters defined as:
\begin{equation}
    \begin{split}
        b_{11} &= \mu_x f_T \hat{\rho}_{\mathrm{eff}} \left(1- f_T \hat{\rho}_{\mathrm{eff}}\right)+ \sigma_x^2 \left(f_T \hat{\rho}_{\mathrm{eff}}\right)^2 \\
    & \qquad + \nu \left( \mu_x f_T \hat{\rho}_{\mathrm{eff}} \epsilon_c \left(1- f_T \hat{\rho}_{\mathrm{eff}} \epsilon_c\right) + \sigma_x^2 \left(f_T \hat{\rho}_{\mathrm{eff}}\epsilon_c\right)^2 \right) \\
    & \qquad + \nu \left(1-\nu\right) \left(1+ \mu_x f_T \hat{\rho}_{\mathrm{eff}}\epsilon_c\right)^2 + \frac{1}{d} \left(1- \frac{1}{d}\right) \\
        b_{12} &= \mu_x f_T \hat{\rho}_{\mathrm{eff}} \left(1- f_T \hat{\rho}_{\mathrm{eff}} \right) + \sigma_x^2 \left (f_T \hat{\rho}_{\mathrm{eff}} \epsilon_c \right)^2  \\
        b_{21} &= \mu_x \left(1-f_T\right) \hat{\rho}_{\mathrm{eff}} \left(1- \left(1-f_T\right)\hat{\rho}_{\mathrm{eff}}\right) + \sigma_x^2 \left(\left(1-f_T\right)\hat{\rho}_{\mathrm{eff}}\right)^2 \\
    & \qquad + \nu \left(\mu_x \left(1-f_T\right)\hat{\rho}_{\mathrm{eff}} \epsilon_c \left(1- \left(1-f_T\right)\hat{\rho}_{\mathrm{eff}}\epsilon_c\right) + \sigma_x^2 \left(\left(1-f_T\right)\hat{\rho}_{\mathrm{eff}}\epsilon_c\right)^2 \right) \\
    & \qquad +\nu \left(1-\nu\right) \left(\mu_x \left(1-f_T\right) \hat{\rho}_{\mathrm{eff}}\epsilon_c \right)^2\\
        b_{22} &= \mu_x \left(1-f_T\right) \hat{\rho}_{\mathrm{eff}} \left(1-\left(1-f_T\right) \hat{\rho}_{\mathrm{eff}}\right) + \sigma_x^2 \left(\left(1-f_T\right)\hat{\rho}_{\mathrm{eff}}\right)^2 + \frac{1}{d}\left(1-\frac{1}{d}\right)
    \end{split}.
\end{equation}
While these expressions are cumbersome, there are two important results to discuss. First, while only the mean number of contacts of an infectious individual ($\mu_x$) appeared in the expression for the mean dynamics (which was then related to $\mathcal{R}_0$), the variance in this distribution $\sigma_x^2$ appears in this set of equations. Second, even though the variance $\sigma_x^2$ appears, it does not affect the stability requirements. Given the structure of this transition matrix, its eigenvalues are less than 1 when the eigenvalues of the matrix $\textbf{A}$ are less than 1. The effective reproduction number $\mathcal{R}_{\mathrm{eff}}$ from Eq.~\ref{eq:ReffStability} remains a sufficient stability criterion for both the mean and variance of the number of infectious individuals. 

\subsection{Simulating infection}
\begin{figure}[htp]
\hspace{-1cm}

\includegraphics[clip,width=0.6\columnwidth]{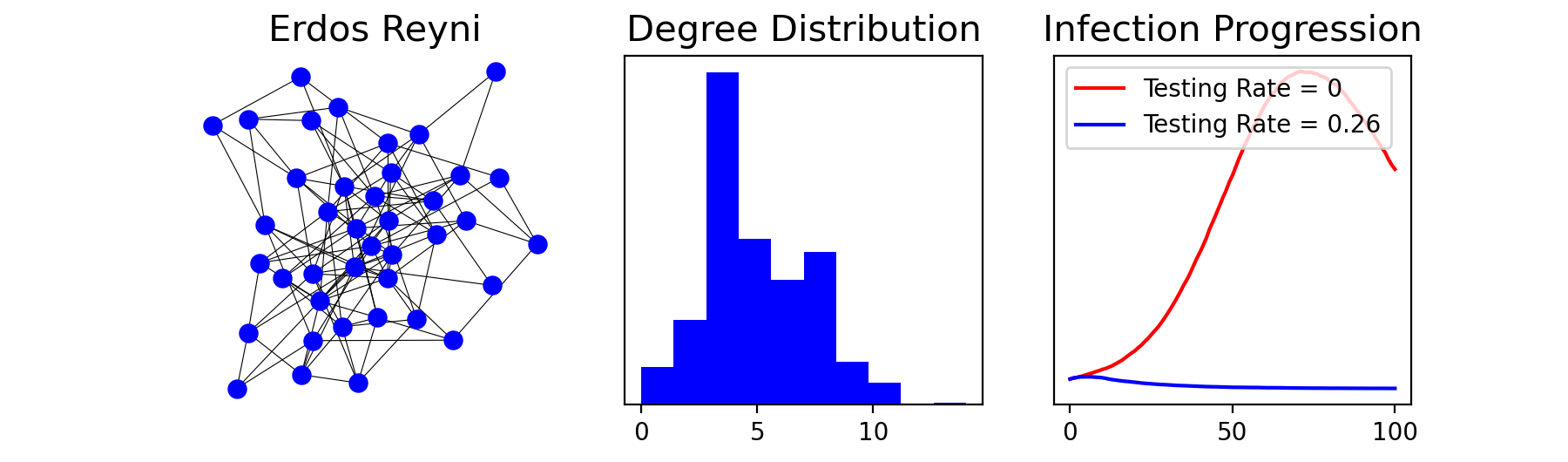}%
\hspace{-1cm}
\includegraphics[clip,width=0.6\columnwidth]{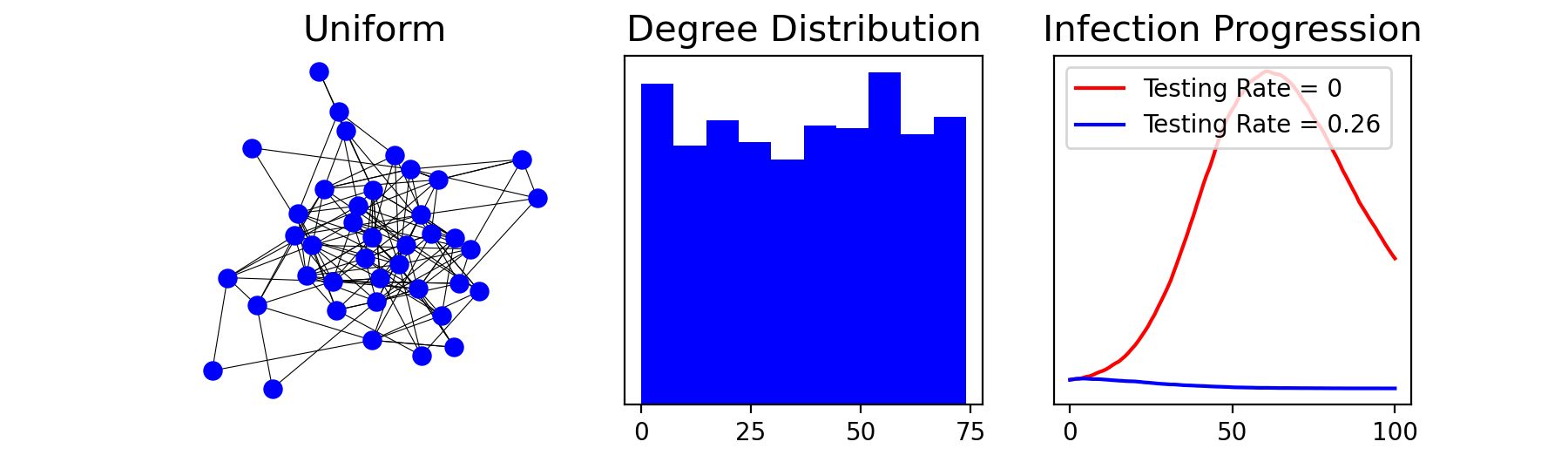}%
\hspace{-1cm}
\includegraphics[clip,width=0.6\columnwidth]{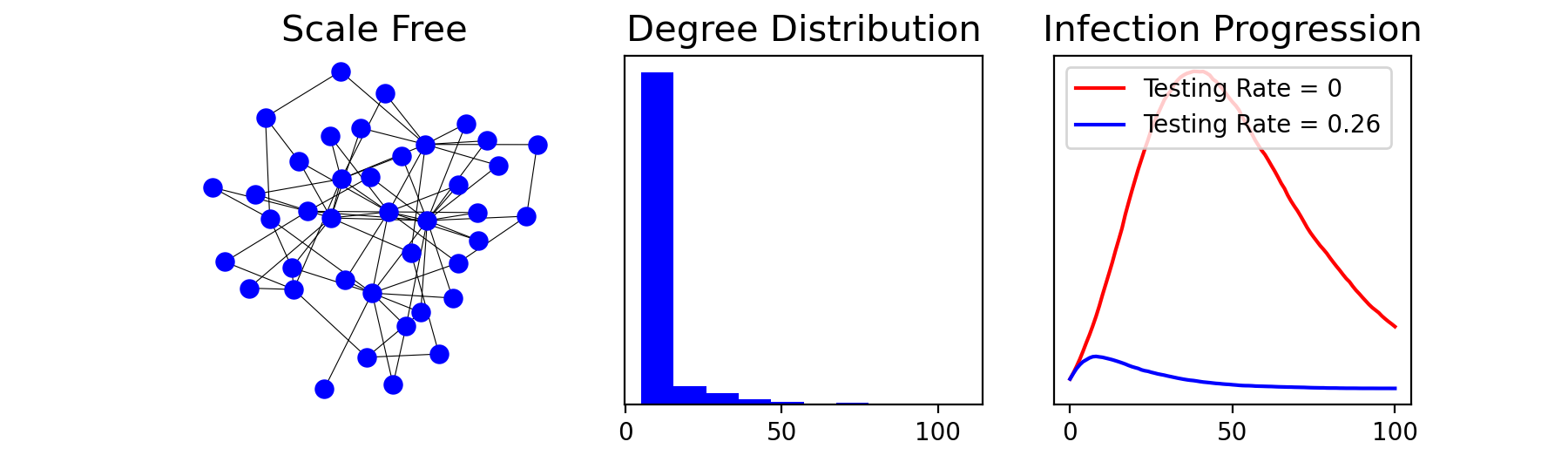}%
\hspace{-1cm}
\includegraphics[clip,width=0.6\columnwidth]{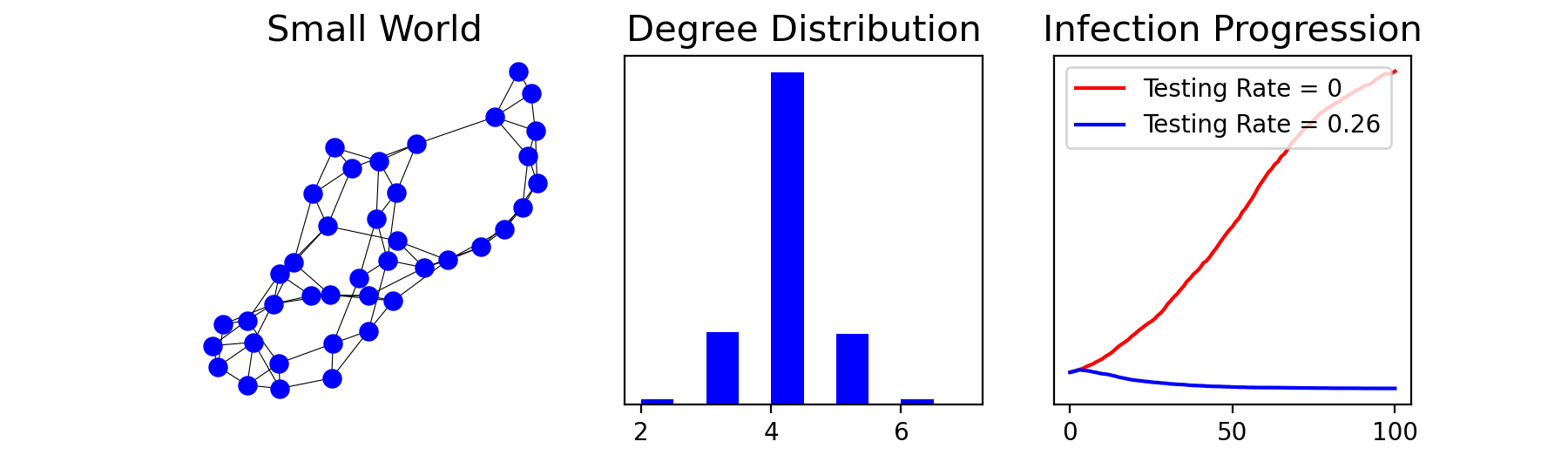}%
\caption{Sample realizations of four different network structures and their resulting degree distributions. Intervention-free infection was simulated on each network to make a measurement of $\mathcal{R}_0$. The required testing to stabilize the system was determined using Eq.~\ref{eq:testingReqFull} and another infection simulation was implemented with testing just above this stabilizing value. Both infection progressions are show: intervention-free (red) and with stabilizing testing (blue).} 
\label{fig:NetworkComparison}
\end{figure}

To test the analytic result derived above in Eq.~\ref{eq:ReffStability}, we implemented an infection simulation on a fixed population. In these simulations, a population network is established at the outset, where every member of the population is a node in the network and every edge represents a contact between two people. Such networks are difficult to estimate for a real population, but generating random networks with wide ranges of structures and properties is straightforward. Fig.~\ref{fig:NetworkComparison} shows random network realizations for four different structure types---Erd\H{o}s-R\'{e}nyi, Uniform, Scale-free, and Small World---and their resulting degree distributions.

Once the underlying network of connections is established, each individual is assigned to one of the partitions of the model as well as to vaccine and testing statuses. At each time step, the infection status of each individual is updated. If a person is infectious, they infect their susceptible neighbors with probability $\rho_{\mathrm{eff}}$ at each time, where $\rho_{\mathrm{eff}}$ is determined by the vaccine and masking status of the individuals involved. If an individual is infectious at time $t$, they recover at $t+1$ with probability $1/d$. The opt-in individuals will become isolated if they are detected through random testing with probability $\nu$. If they are detected, each of their infectious neighbors has a probability $\epsilon_c$ of also joining the isolated population. Isolated individuals will eventually recover and do so with probability $1/d$ at each time step. The simulation continues until there are no longer any infectious individuals remaining or for a fixed number of days.

\section{RESULTS}
\subsection{Balancing mitigation strategies: analysis}
Having performed the stability analysis, we turn our attention to the practical implications of the effective reproduction number of Eq.~\ref{eq:ReffStability}. The stability criterion ($\mathcal{R}_{\mathrm{eff}} < 1$) can be rearranged to provide guidance on how much of each mitigation strategy an organization should employ to ensure their reopening does not lead to widespread outbreaks. 

\subsubsection{Without testing}
Let us first consider the case where no testing is done $\nu = 0$, which is also equivalent to the case where testing is available (nonzero $\nu$) but no one opts into the program ($f_T= 1$). This will be the case for the many organizations that do not have access to testing resources or the funding to secure them. It would also be the case for organizations where the vast majority of individuals are opposed to being testing and a testing requirement cannot be implemented. In this case, the stability criterion simplifies to
\begin{equation}
    1 > \mathcal{R}_0 \left(1-\epsilon_m f_m\right)^2 \left(1- \epsilon_v f_v\right) = \mathcal{R}_{\nu = 0}  \label{eq:noTesting}
\end{equation}
In the absence of masking and vaccinations, this simplifies to the well-known result for stable dynamics in the mitigation-free case: $1 > \mathcal{R}_0$. Implementing masking and/or vaccination requirements, however, will allow for controlled dynamics even in the case of basic reproduction numbers larger than 1. 

First let us consider the effect of masking in the absence of vaccines, as would be the case at the beginning of a disease epidemic when vaccines may not exist or may not be widely available. Given a disease ($\mathcal{R}_0$) and the efficacy of masks at preventing transmission of that disease $\epsilon_m$, we can determine the fraction of the population that must be masked in order to ensure stability to disease outbreaks:
\begin{equation}
    f_m > \frac{1- \frac{1}{\sqrt{\mathcal{R}_0}}}{\epsilon_m}.
\end{equation}
We are of course limited by the fact that no more than 100\% of the population can be masked at a given time. Meaning if $\mathcal{R}_0 > 1/\left(1-\epsilon_m\right)^2$, the disease cannot be controlled by masking alone.

Similar analysis can be performed to determine the effect of vaccination on the spread of a disease. Because compliance to masking can be difficult to ensure and because masking can be unpleasant and a hindrance to some activities, we now aim to determine what fraction of the population needs to be vaccinated to ensure that, in the absence of testing, no masking is required. The required vaccinated fraction is 
\begin{equation}
    f_v > \frac{1- \frac{1}{\mathcal{R}_0}}{\epsilon_v}.
    \label{eq:vaxMandate}
\end{equation}
This recovers the classical definition of \textit{herd immunity}, i.e.~the fraction of the population that needs to be vaccinated to control the spread of a disease. Similar to the analysis of masking, we are limited by the fact that no more than 100\% of the population can be vaccinated, the result being that only diseases with $R_0 < 1/\left(1-\epsilon_v\right)$ can be controlled by vaccine mandates alone. In reality, less than 100\% of the population will be vaccinated, and if Eq.~\ref{eq:vaxMandate} does not hold for that value of $f_v$, additional mitigation strategies will be required to ensure stable disease dynamics. Such additional strategies may include masking (and the combined effects of these two interventions together is captured in Eq.~\ref{eq:noTesting}) or testing and contact tracing, as discussed below.

\subsubsection{With testing}
Now we will consider the case where testing resources are available to a given organization. We will first assume that participation in the testing program is mandatory for all individuals, i.e.~$f_T = 1$. In this case, the stability criterion of Eq.~\ref{eq:ReffStability} simplifies to 
\begin{equation}
     1 > \mathcal{R}_{\nu=0} \left(\frac{1-\nu \epsilon_c}{1 + \nu \left(d-1\right)}\right). 
\end{equation}
where $\mathcal{R}_{\nu=0}$ is the effective reproduction number in the absence of testing but includes the effects of masking and vaccinations (see Eq.~\ref{eq:noTesting}). This expression can be rearranged to determine the minimum required daily testing fraction
\begin{equation}
    \nu_{f_T = 1} > \frac{\mathcal{R}_{\nu=0} -1}{\epsilon_c \mathcal{R}_{\nu=0} + d - 1}.
    \label{eq:testingMin}
\end{equation}
Again, this expression holds for the case where everyone opts into the testing program. We also recover the fact that if the masking and vaccine interventions can drive the effective reproduction number $\mathcal{R}_{ \nu=0}$ below 1, then no testing is required ($\nu_{f_T=1} = 0$). We further recover the result that if $R_0 < 1$, even without masking and vaccines, no testing is required to ensure stability to disease outbreaks. 

Unlike the masking and vaccinated fractions ($f_m$ and $f_v$), the daily testing fraction can exceed $1$, meaning individuals are tested (on average) more than once per day. In reality, this is unlikely due to available testing resources. Additionally, rather than random testing, organizations may prefer to perform fixed-cadence testing for simpler logistical implementation. For instance, organizations may choose to test once a day (with an effective testing fraction $\nu=1$), once a week ($\nu = 1/7$), or once every two weeks ($\nu = 1/14$). Even if the testing capability is larger than the given effective testing fraction, organizations may still opt to test on a regular cadence to keep the implementation as easy as possible. 

Finally, before discussing the case where individuals only participate in the testing program voluntarily ($f_T < 1$), the effect of contact tracing must be noted. By examining Eq.~\ref{eq:testingMin}, we see that less contact tracing (lower efficacy $\epsilon_c$) corresponds to increased required daily testing fractions and thus increased required daily testing resources. Organizations with the capability to do contact tracing (and to do so at little to no cost) should therefore employ this strategy to reduce the burden and cost of testing. 

Next we consider the case where not all individuals are required to participate in surveillance testing (or if they are required to but are not compliant). Eq.~\ref{eq:ReffStability} can again be rearranged to provide guidance on how much testing must be done in this case:
\begin{equation}
    \nu > \frac{\mathcal{R}_{\nu=0} -1}{\left[f_T\epsilon_c - \left(1-f_T\right) \left(d-1\right)\right]\mathcal{R}_{\nu=0} + d - 1 }
    \label{eq:testingReqFull}
\end{equation}
where again, $\mathcal{R}_{\nu=0}$ is the effective reproduction number in the presence of masking and vaccinations but the absence of testing (see Eq.~\ref{eq:noTesting}). This expression captures the effect of all of the mitigation strategies we have explored. Given the masking ($f_m$), vaccination ($f_v$), and testing ($f_T$) requirements of an organization, this relationship determines the amount of daily testing that must be done ($\nu$) to ensure reopening without risk of causing widespread outbreaks of a specific disease ($\mathcal{R}_0, d$). If an organization does not have sufficient resources to perform this level of daily testing, they can alter their requirements for the other mitigation strategies to lower the needed testing. 
\subsection{Balancing mitigation strategies: example}
While the trade-offs between the discussed mitigation strategies are fully expressed in Eq.~\ref{eq:testingReqFull}, they are nonlinear and perhaps difficult to imagine. In this section, an illustration of these trade-offs is provided for a specific set of disease parameters. These parameters loosely correspond to the reported values for the Delta variant of SARS-CoV-2. The selected basic reproduction number is $\mathcal{R}_0 = 5$  \cite{deltaR0_5}; the length of the contagious period is $d=14$ days \cite{d_14}; and the efficacies of masking and vaccinations are $\epsilon_m = 0.25$ \cite{hosoi2021estimating} and $\epsilon_v = 0.65$ \cite{vaccineEfficacy}. 

For this particular disease, we consider four scenarios: no masking and no vaccines ($f_m = f_v = 0$), only masking with full compliance ($f_m = 1, f_v = 0$), only vaccinating with full compliance ($f_m = 0, f_v =1$), and full compliance on both masking and vaccination mandates ($f_m=f_v=1$). In each scenario, we consider the full spectrum of compliance to participation in a testing program ($f_T = [0,1]$), and testing scenarios from none at all to daily testing for everyone ($\nu = [0,1]$). In all scenarios, contacts of infectious individuals are assumed to be successfully identified and tracked down with efficacy $\epsilon_c = 0.8$. 

\begin{figure}
\centering
\includegraphics[trim = 100 50 100 0, clip, width=1\columnwidth]{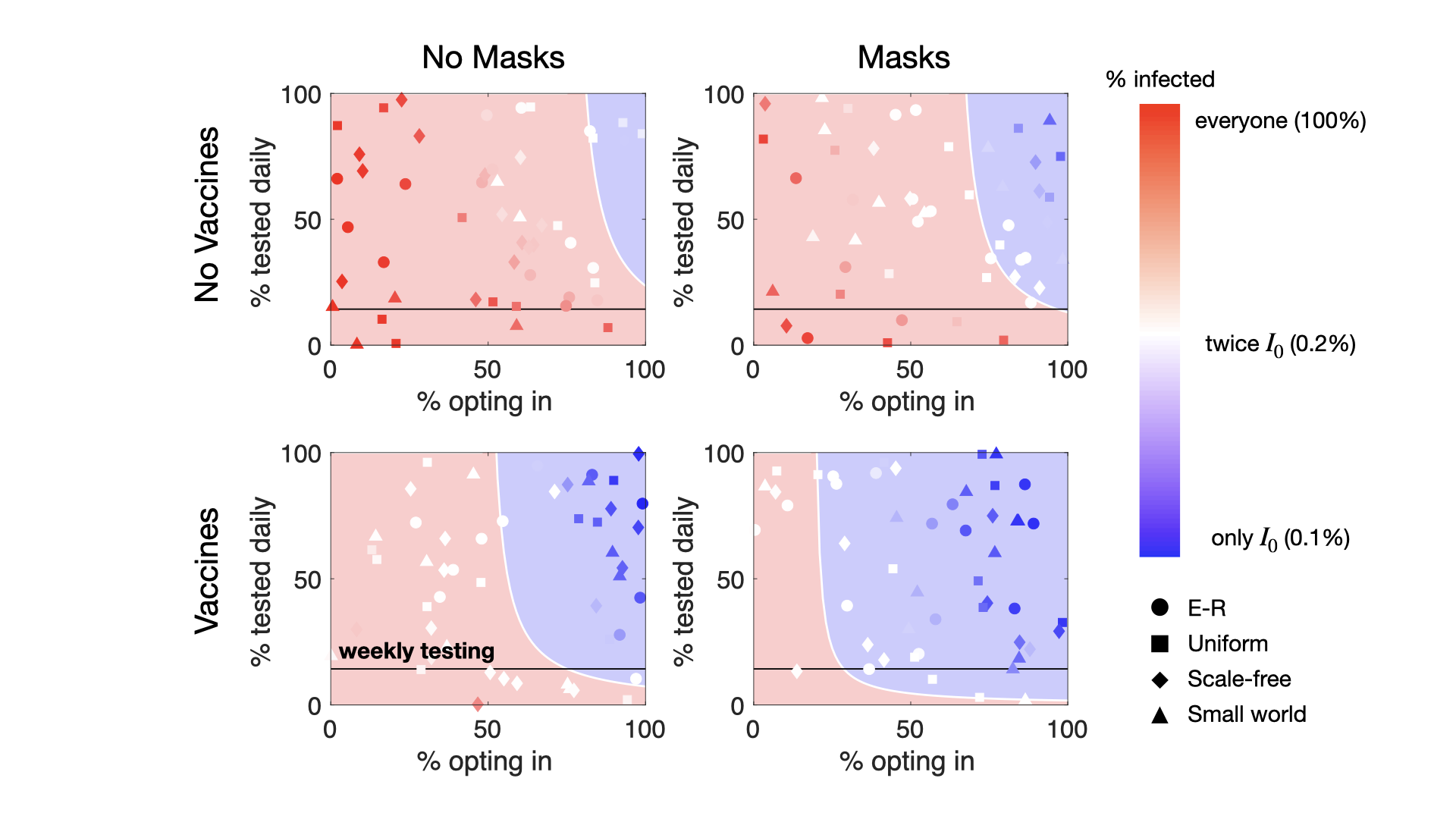}
\caption{Trade-offs between mitigation strategies. Infection of a disease similar to the Delta variant of SARS-CoV-2 with $\mathcal{R}_0 = 5$ and $d = 14$ was simulated on random networks of various structures under random intervention combinations. The analytical stability boundary is shown as a white line separating the controlled region (blue background) from the uncontrolled region (red background). Points represent the average over 100 infection simulations for the same set of parameters on a random network whose structure is represented by point shape.}
\label{fig:tradeoffs}
\end{figure}

Using these values and Eq.~\ref{eq:testingReqFull}, a stability boundary was determined in the space spanning the testing frequency ($\nu$) and testing compliance ($f_T$) for each of these scenarios. This stability boundary is shown for each of the four scenarios as white lines in Fig.~\ref{fig:tradeoffs}. If the combination of daily testing fraction and fraction of the population  opting-in to the testing program falls above the stability boundary (in a blue region in Fig.~\ref{fig:tradeoffs}), the mitigation strategies in place will lead to successful prevention of widespread outbreaks of this disease. If instead a community's testing program parameters fall below the stability boundary (in a red region in Fig.~\ref{fig:tradeoffs}), that community would be susceptible to outbreaks. 

To further illustrate the separation of stable and unstable dynamics, the propagation of a disease with these parameters was simulated on 200 random networks (50 for each scenario) comprised of 5000 nodes with a randomly selected testing program ($f_T, \nu$ combination). The structure and properties of these networks were selected randomly from the ranges specified in Table \ref{table:MC}. In Fig.~\ref{fig:tradeoffs}, each simulated network is represented as a point whose shape indicates its network structure and whose color represents the percent of the population that became infected with the disease over 180 days (averaged over 100 random infection progressions for initial seeded infections of $I_0 = 5$). Notice, the color scale is nonlinear; blue-to-white points represent those where each initially infected individual ($5$ people) spread the disease to, on average, fewer than one other individual ($<10$ people infected total), while white-to-red points represent those where the disease spread more than that, up to propagating through the entire population (5000 people). The simulation results are consistent with the stability analysis performed above. 

The results in Fig.~\ref{fig:tradeoffs} are an illustration of the analysis presented in this paper only for one specific disease (based on the Delta variant of SARS-CoV-2), but the analysis will hold for diseases of all parameters. For instance, in the case of both mask and vaccine mandates (the lower right plot of Fig.~\ref{fig:tradeoffs}), some amount of regular testing and compliance to testing is required to ensure a community will be stable to widespread outbreaks of this disease. However, if a different set of disease parameters were selected, e.g.~ones based on the Alpha variant of SARS-CoV-2 which will have a lower basic reproduction number and against which vaccines will be more effective \cite{vaccineEfficacy, d_14}, mask and vaccine mandates would be sufficient for controlling the spread of the disease (see Eq.~\ref{eq:noTesting}), and testing would no longer be needed. Finally, while Fig.~\ref{fig:tradeoffs} only shows four scenarios based on the absence/presence of fully compliant masking and/or vaccine mandate, compliance to either mandate in practice would likely fall somewhere in between. The stability boundary in these scenarios can still be calculated using Eq.~\ref{eq:testingReqFull} and used to determine the requirements of a successful testing program. 

\subsection{Network (in)dependence}
The simulations whose results are shown in Fig.~\ref{fig:NetworkComparison} and Fig.~\ref{fig:tradeoffs} both begin to suggest the minimal role of the network parameters on the stability criterion. All relevant network information is captured in the value of $\mathcal{R}_{0}$; no additional information about the network structure or degree distribution is required. This result is clear from the expression for the effective reproduction number (Eq.~\ref{eq:ReffStability}) but is further illustrated by performing a Monte Carlo simulation of the parameter space. 
\begin{figure}
\includegraphics[trim = 0 125 0 0, clip, width=0.8\columnwidth]{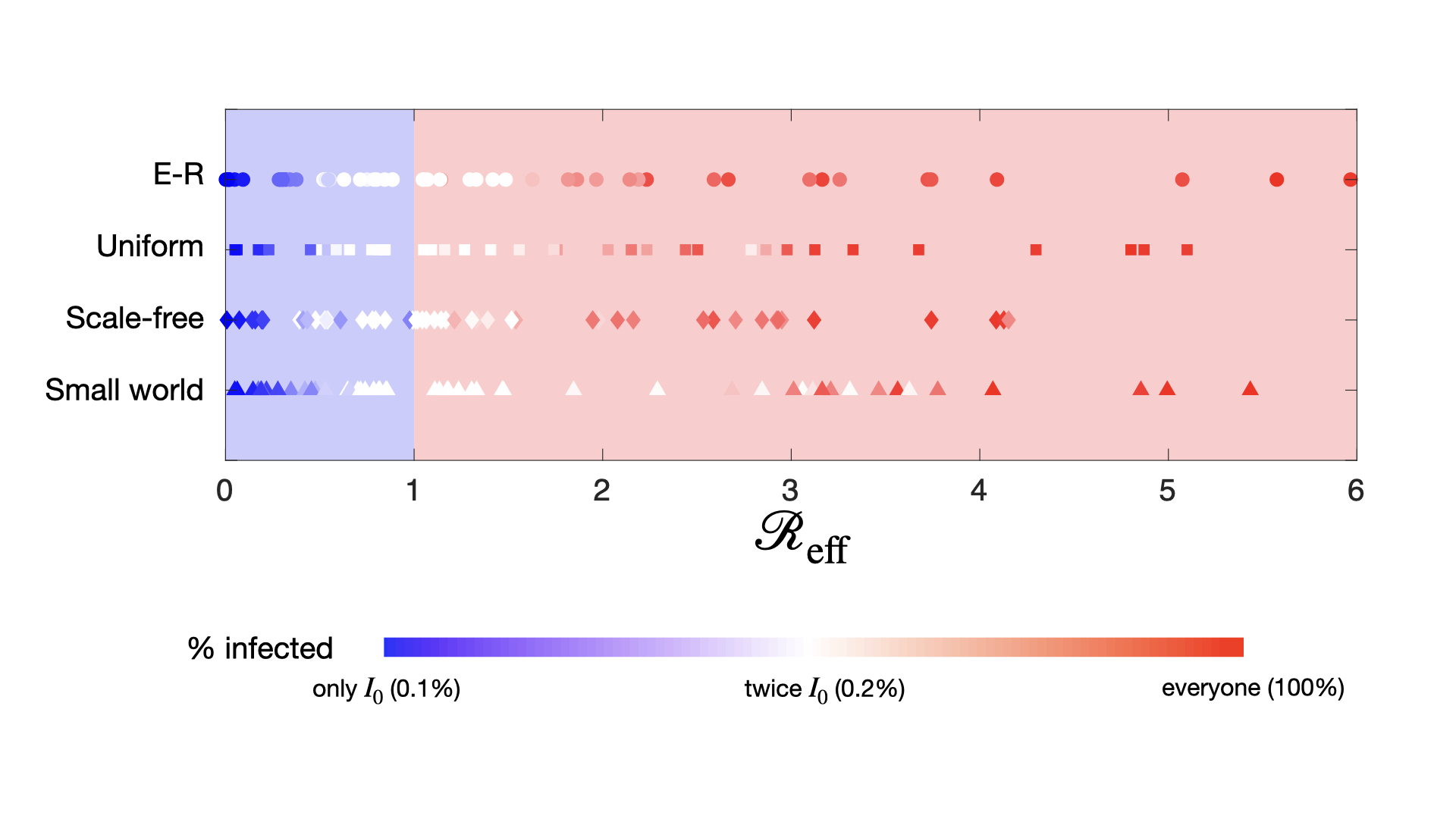}
\caption{Monte Carlo infection simulation. Intervention strategies, intervention efficacies, disease parameters, and network parameters were all generated randomly. The mean total fraction of the population which became infected over 100 trials at each combination of parameters is shown (via point color) against the effective reproduction number $\mathcal{R}_{\mathrm{eff}}$. This single parameter successfully separates the controlled and uncontrolled infection outbreaks denoted by blue and red backgrounds respectively.}
\label{fig:oneDplot}
\end{figure}

The results of this Monte Carlo simulation are shown in Fig.~\ref{fig:oneDplot} and the ranges from which all the parameters were sampled are listed in Table \ref{table:MC}. Networks of 5000 nodes were generated and seeded randomly with 5 initial infections. Fig.~\ref{fig:oneDplot} presents the fraction of the population that became infected over the whole simulated epidemic (on average over 100 random infection progressions) against the $\mathcal{R}_{\mathrm{eff}}$ (computed from the randomly selected combination of parameters using Eq.~\ref{eq:ReffStability}). For all combinations and networks, only this effective reproduction number is required to predict whether outbreaks are limited to $<1\%$ of the population or spread to nearly everyone. While the exact number of individuals who become infected may vary, the presence or absence of an outbreak is well predicted by the effective reproduction number, regardless of the network structure. Again, this is consistent with the analysis in which $\mathcal{R}_0$ captures all of the relevant network information; no further information about the distribution of edges (contacts) in the network, e.g.~the mean or variance of this distribution, is required to determine what combination of mitigation strategies will successfully lead to controlled disease dynamics.
\begin{table}[htbp]
\caption{The ranges for each parameter that was sampled randomly and used in an infection simulation.}
\begin{center}
\begin{tabular}{l c|c|c}
\multicolumn{2}{c|}{\textbf{Parameter}}&\textbf{Description}&\textbf{Range}\\
\hline{}
Disease & $\mathcal{R}_0$ & basic reproduction number & 3--13 \\
& $d$ & contagious period (days) & 5--15 \\
\hline
Intervention & $\epsilon_m$ & mask efficacy & 0--1 \\
& $\epsilon_v$ & vaccine efficacy & 0--1 \\
& $\epsilon_c$ & contact tracing efficacy & 0--1 \\
& $f_m$ & fraction masked & 0--1 \\
& $f_v$ & fraction vaccinated & 0--1 \\
& $f_T$ & fraction opting-in to testing & 0--1 \\
& $\nu$ & daily testing fraction & 0--1 \\
\hline
Network & & & \\
\textit{Erd\H{o}s-R\'{e}nyi} & $\mu_{\mathrm{ER}}$ & mean degree & 5--25 \\
\textit{Uniform} & $x_{\mathrm{min,U}}$ & minimum degree & 1--10 \\
& $x_{\mathrm{max,U}}$ & maximum degree & 10--20 \\
\textit{Scale-free} & $\alpha_{\mathrm{SF}}$ & distribution power law & 3--5 \\
& $x_{\mathrm{min,SF}}$ & minimum degree & 1--20 \\
\textit{Small World} & $\mu_{\mathrm{SW}}$ & mean degree & 5--25 \\
& $\beta_{\mathrm{SW}}$ & rewiring probability & 0--1 
\end{tabular}
\label{table:MC}
\end{center}
\end{table}

\section{DISCUSSION}
\subsection{Comparison to university case data}
This analysis was performed to inform policy making for semi-contained organizations with a limited number of individuals, universities in particular. Since the large-scale shutdowns of the early COVID-19 pandemic, most universities have returned to nearly full operation using some combination of the mitigation strategies discussed above. Few universities have reported widespread outbreaks since their reopenings. This is consistent with our analysis and may be because schools are implementing a sufficient combination of mitigation strategies to prevent such outbreaks. However, a lack of reported outbreaks does not necessarily imply stability; universities may not be reporting outbreaks because they are simply not detecting them. Detection requires regular viral testing of some kind. If schools are testing regularly, they would detect outbreaks. However, testing weekly would also be a sufficient way to ensure that there are no outbreaks.\footnote{The Delta variant of SARS-CoV-2 was discovered after the vaccine was developed, and as seen in Fig.~\ref{fig:tradeoffs}, weekly testing in the case of a highly vaccinated population is sufficient for controlling the spread of this disease regardless of masking behavior. In reality, compliance to vaccine requirements will be less that $100\%$, but weekly testing is likely still sufficient provided enough individuals opt-into the program. Therefore, schools that are testing their members once a week are implementing mitigation strategies that our analysis suggests are sufficient for preventing outbreaks.} Universities that are not testing may not be in the stable disease dynamics regime but may also not detect instability.

Case data can also be misleading as cases can go up at a university even if appropriate mitigation strategies are in place. This happens because each day there can be interactions between those inside the community (e.g.~students at the school) and those outside the community (e.g.~individuals in the town or city in which the school is located). These interactions effectively lead to new ``seed" infections in the community population. While the policies in place at the university can control whether or not these seeds grow, they may be limited in the control of how often the seeds are planted; the seeding rate will depend more on the overall prevalence of the disease in the area. Separating case data at a university by the origin of the infection (from intra-community or extra-community interactions) is a complex data task beyond the scope of this analysis.

\subsection{Generality of results}
While this work was originally inspired by the necessity to use testing and other interventions to control the spread of COVID-19, the results depend neither on the details of the disease nor on the details of the population structure. Given some estimate of the parameters of a disease's proliferation, the stability analysis performed in this paper gives general guidelines on what successful testing programs require based on the adherence to masking and/or vaccination policies by individuals in the population. 

The resulting stability criterion interestingly \emph{does not} depend on the variance of the underlying network of the community. Highly connected individuals only impact the dynamics through their contribution to the measured basic reproduction number $\mathcal{R}_0$. This simplifies the analysis of the spread of an epidemic as all the relevant network details are captured in the measured $\mathcal{R}_0$, allowing one to evaluate mitigation strategies for all communities, not just those whose interaction behaviors are roughly homogeneous. Hence in examining the impact of interventions on \textit{stability}, higher complexity network models (those with additional information about the underlying community structure) are unnecessary. Analyses that go beyond stability, e.g.~predicting the exact size of an outbreak rather than whether or not an outbreak will exist, require models that contain more detailed network information \cite{SakaguchiInhomogeneous,RadicchiOutbreakSize}. However, for the prevention of widespread outbreaks, the stability criterion presented in Eq.~\ref{eq:ReffStability} is sufficient for capturing the effect of different mitigation strategies in a way that allows policy makers to balance trade-offs, given their available resources, how to reopen their schools, businesses, and other organization safely amidst a pandemic.

\bibliographystyle{plain}
%
\bibliography{pandemicControl}

\begin{thebibliography}{10}

\bibitem{berger2020seir}
David~W. Berger, Kyle~F. Herkenhoff, and Simon Mongey.
\newblock An {SEIR} infectious disease model with testing and conditional
  quarantine.
\newblock {\em NBER Working Papers}, page~1, 2020.

\bibitem{DelamaterPaulL2019ComplexR0}
Paul~L Delamater, Erica~J Street, Timothy~F Leslie, Y.~Tony Yang, and Kathryn~H
  Jacobsen.
\newblock Complexity of the basic reproduction number ({R0}).
\newblock {\em Emerging infectious diseases}, 25(1):1--4, 2019.

\bibitem{lonely1}
Roni Elran-Barak and Maya Mozeikov.
\newblock One month into the reinforcement of social distancing due to the
  {COVID-19} outbreak: Subjective health, health behaviors, and loneliness
  among people with chronic medical conditions.
\newblock {\em International journal of environmental research and public
  health}, 17(15), 2020.

\bibitem{SIQRmodel}
Z~Feng and H~R Thieme.
\newblock Recurrent outbreaks of childhood diseases revisited: the impact of
  isolation.
\newblock {\em Mathematical biosciences}, 128(1-2):93 -- 130, 1995.

\bibitem{ageModel}
Daniel Franco, Hartmut Logemann, and Juan Perán.
\newblock Global stability of an age-structured population model.
\newblock {\em Systems \& Control Letters}, 65:30 -- 36, 2014.

\bibitem{asympSpread}
Daihai He, Shi Zhao, Qianying Lin, Zian Zhuang, Peihua Cao, Maggie~H. Wang, and
  Lin Yang.
\newblock The relative transmissibility of asymptomatic {COVID-19} infections
  among close contacts.
\newblock {\em International Journal of Infectious Diseases}, 94:145 -- 147,
  2020.

\bibitem{HeffernanJM2005_measureR0}
J.M Heffernan, R.J Smith, and L.M Wahl.
\newblock Perspectives on the basic reproductive ratio.
\newblock {\em Journal of the Royal Society interface}, 2(4):281--293, 2005.

\bibitem{d_14}
Gabriel~G. Katul, Assaad Mrad, Sara Bonetti, Gabriele Manoli, and Anthony~J.
  Parolari.
\newblock Global convergence of covid-19 basic reproduction number and
  estimation from early-time sir dynamics.
\newblock {\em PLOS ONE}, 15(9):1--22, 09 2020.

\bibitem{KEELINGMATTJ2000_measureR0}
Matt~J Keeling and Bryan~T Grenfell.
\newblock Individual-based perspectives on {R0}.
\newblock {\em Journal of theoretical biology}, 203(1):51--61, 2000.

\bibitem{SIR_OG}
WO~Kermak and AG~McKendrick.
\newblock Contributions to the mathematical-theory of epidemics 1. (reprinted
  from proceedings of the royal society, vol 115a, pg 700-721, 1927).
\newblock {\em Bulletin of Mathematical Biology}, 53(1-2):33 -- 55, 1991.

\bibitem{KojakuSadamori2021_friendship}
Sadamori Kojaku, Laurent Hébert-Dufresne, Enys Mones, Sune Lehmann, and
  Yong-Yeol Ahn.
\newblock The effectiveness of backward contact tracing in networks.
\newblock {\em Nature physics}, 17(5):652--658, 2021.

\bibitem{COVIDAsymptomatic}
Andreas Kronbichler, Daniela Kresse, Sojung Yoon, Keum~Hwa Lee, Maria
  Effenberger, and Jae~Il Shin.
\newblock Asymptomatic patients as a source of {COVID-19} infections: A
  systematic review and meta-analysis.
\newblock {\em International Journal of Infectious Diseases}, 98:180 -- 186,
  2020.

\bibitem{LiJing2011T_failureR0}
Jing Li, Daniel Blakeley, and Robert~J Smith.
\newblock The failure of {R0}.
\newblock {\em Computational and mathematical methods in medicine}, 2011, 2011.

\bibitem{deltaR0_5}
Ying Liu and Joacim Rocklöv.
\newblock {The reproductive number of the Delta variant of SARS-CoV-2 is far
  higher compared to the ancestral SARS-CoV-2 virus}.
\newblock {\em Journal of Travel Medicine}, 28(7), 08 2021.
\newblock taab124.

\bibitem{vaccineEfficacy}
Jamie Lopez~Bernal, Nick Andrews, Charlotte Gower, Eileen Gallagher, Ruth
  Simmons, Simon Thelwall, Julia Stowe, Elise Tessier, Natalie Groves, Gavin
  Dabrera, Richard Myers, Colin~N.J. Campbell, Gayatri Amirthalingam, Matt
  Edmunds, Maria Zambon, Kevin~E. Brown, Susan Hopkins, Meera Chand, and Mary
  Ramsay.
\newblock Effectiveness of covid-19 vaccines against the b.1.617.2 (delta)
  variant.
\newblock {\em New England Journal of Medicine}, 385(7):585--594, 2021.
\newblock PMID: 34289274.

\bibitem{lonely2}
José~G. Luiggi-Hernández and Andrés~I. Rivera-Amador.
\newblock Reconceptualizing social distancing: Teletherapy and social
  inequality during the {COVID-19} and loneliness pandemics.
\newblock {\em Journal of Humanistic Psychology}, 60(5):626 -- 638, 2020.

\bibitem{hosoi2021estimating}
Xinyu Mao and AE~Hosoi.
\newblock Estimating the filtration efficacy of cloth masks.
\newblock {\em Physical Review Fluids}, 6(11):114201, 2021.

\bibitem{IntroMathEpid}
M.~Martcheva.
\newblock {\em Introduction to mathematical epidemiology.}
\newblock Texts in applied mathematics: v. 61. New York : Springer, [2015],
  2015.

\bibitem{socialDistancing1}
Daniel~J. McGrail, Jianli Dai, Kathleen~M. McAndrews, and Raghu Kalluri.
\newblock Enacting national social distancing policies corresponds with
  dramatic reduction in {COVID-19} infection rates.
\newblock {\em PLoS ONE}, 15(7):1 -- 9, 2020.

\bibitem{econ1}
Fabio Milani.
\newblock {COVID-19} outbreak, social response, and early economic effects: a
  global {VAR} analysis of cross-country interdependencies.
\newblock {\em Journal of population economics}, pages 1 -- 30, 2020.

\bibitem{RadicchiOutbreakSize}
Filippo Radicchi and Ginestra Bianconi.
\newblock Epidemic plateau in critical susceptible-infected-removed dynamics
  with nontrivial initial conditions.
\newblock {\em Phys. Rev. E}, 102:052309, Nov 2020.

\bibitem{SakaguchiInhomogeneous}
Hidetsugu Sakaguchi and Yuta Nakao.
\newblock Slow decay of infection in the inhomogeneous
  susceptible-infected-recovered model.
\newblock {\em Phys. Rev. E}, 103:012301, Jan 2021.

\bibitem{econ2}
Heli Simola and Laura Solanko.
\newblock Domestic and global economic effects of {COVID-19} containment
  measures: How does domestic and global economic effects of {COVID-19}
  containment measures: How does {Russia} compare internationally?.
\newblock {\em BOFIT Policy Brief}, (6):3 -- 10, 2020.

\bibitem{socialDistancing3}
Alexandra Teslya, Thi~Mui Pham, Noortje~G. Godijk, Mirjam~E. Kretzschmar,
  Martin C.~J. Bootsma, and Ganna Rozhnova.
\newblock Impact of self-imposed prevention measures and short-term
  government-imposed social distancing on mitigating and delaying a {COVID-19}
  epidemic: A modelling study.
\newblock {\em PLoS Medicine}, 17(7):1 -- 21, 2020.

\bibitem{socialDistancing2}
Tran Phuoc~Bao Thu, Pham Nguyen~Hong Ngoc, Nguyen~Minh Hai, and Le~Anh Tuan.
\newblock Effect of the social distancing measures on the spread of {COVID-19}
  in 10 highly infected countries.
\newblock {\em Science of the Total Environment}, 742, 2020.

\bibitem{CDCCOVID}
{United States Centers for Disease Control and Prevention}.
\newblock {\em {United States COVID-19 Cases, Deaths, and Laboratory Testing
  (NAATs) by State, Territory, and Jurisdiction}}, 2021 (accessed November 29,
  2021).

\bibitem{SEIRModel2}
Juan Zhang and Zhien Ma.
\newblock Global dynamics of an {SEIR} epidemic model with saturating contact
  rate.
\newblock {\em Mathematical Biosciences}, 185(1):15 -- 32, 2003.

\end{thebibliography}

\end{document}